\input{aipcheck}

\documentclass[
    ,final            
  ]
  {aipproc}

\layoutstyle{6x9}

\begin{document}

\title{White Dwarf Planets from GAIA}



\author{Roberto Silvotti}{address={INAF-Osservatorio Astronomico di Torino, strada dell'Osservatorio 20, 10025 Pino Torinese, Italy}}

\author{Alessandro Sozzetti}{address={INAF-Osservatorio Astronomico di Torino, strada dell'Osservatorio 20, 10025 Pino Torinese, Italy}}

\author{Mario Lattanzi}{address={INAF-Osservatorio Astronomico di Torino, strada dell'Osservatorio 20, 10025 Pino Torinese, Italy}}

\begin{abstract}
We investigate the potential of high-precision astrometry with GAIA for detection of giant planetary companions to nearby white dwarfs. 
If one considers that, to date, no confirmed planets around single white dwarfs are known, the results from GAIA will be crucial to study the late-stage evolution of planetary systems and to verify the possibility that 2nd-generation planets are formed.
\end{abstract}

\maketitle


\section{Introduction}

The evolution of the planetary systems after the main sequence (MS) and after the red giant branch (RGB) is largely unknown. 
The possibility to detect and study planets around white dwarfs (WDs) represents a major step for our understanding of the late-stage evolution of the planetary systems and can give us the ability to trace back their entire history from the
protoplanetary disk to the MS and, through different evolutionnary paths, to the final configuration when the star becomes a white dwarf.

The first discoveries of planets orbiting post-RGB stars are very recent 
(see Silvotti et al. and other articles of these proceedings) and very recent are also the first theoretical models to study the star-planet interaction during the most critical phases of the stellar evolution: RGB expansion, asymptotic giant branch expansion and planetary nebula ejection (e.g. Villaver, these proc.).

However, until now, we know only a few planets around wide MS+WD binaries
(HD\,13445=Gl\,86, HD\,27442 and HD\,147513, Desidera \& Barbieri 2007 and ref. therein)
and a planet around the pulsar+WD system PSR 1620-26 (Thorsett et al. 1999, 
Sigurdsson et al. 2003),
while there are no clear detections of planets around single white dwarfs, with only a few candidates (GD\,66, Mullally et al. 2008, 2009; GD\,356, Wickramasinghe et al. 2010). 

%
%
%
%
%

\section{The potential of GAIA to detect WD planets}

Among the $\approx$400,000 white dwarfs that will be detected by GAIA 
\cite{Jordan2007}, for the most close and bright objects it will be possible 
to check the presence of low-mass companions, down to brown dwarfs and planets.
In order to evaluate the number of these targets, we use the WD local sample 
within 20 pc (Holberg et al. 2008, Sion et al. 2009) to determine the WD space 
density in each bin of absolute V magnitude. 
According to \cite{Holberg+2008}, the sample is considered to be 80\% complete 
and the densities are corrected for incompleteness. 
For the hottest objects with absolute magnitude 8$<$V$_{abs}$$<$11, the 
statistics of the WD local sample is too poor and hence we use instead a 
theoretical estimate of the space density, proportional to the evolutionary 
duration of each bin, following Bergeron's DA WD models (Holberg \& 
Bergeron 2006, see also 
{\it http://www.astro.umontreal.ca/~bergeron/CoolingModels}). 
For the coolest objects with V$_{abs}$$>$15, again the statistics of the 
WD local sample is poor and probably the completeness is less than 80\% but 
in any case these faint stars have little influence on our counts as, for them,
the volume sampled is small.

Table 1 reports the space densities (column 2) and the number of white dwarfs 
within 50 pc (with apparent magnitude V$\le$13) or 100 pc (and V$\le$15) in the 
last 2 columns. 
In column 3-4 and 5-6 we give also the distance at which we reach the V
magnitude 13 or 15 and the corresponding volumes.

Figure 1 shows the mass of the WD planets detectable with GAIA vs. their orbital distance.
In order to contextualize the potential of GAIA, Fig.1 reports also the known exoplanets detected with various methods (radial velocities, transits, microlensing and timing) and a theoretical planet distribution from Ida \& Lin 
(2008). 

\begin{table}
\begin{tabular}{ccrrrrrrr}
\hline
    \tablehead{1}{c}{b}{V$_{abs}$\\}
  & \tablehead{1}{c}{b}{WD space\\density [pc$^{-3}$]}
  & \tablehead{1}{c}{b}{d$_{13}$~[pc]\tablenote{distance corresponding to an apparent magnitude V=13}\\}
  & \tablehead{1}{c}{b}{d$_{15}$~[pc]\tablenote{distance corresponding to an apparent magnitude V=15}\\}
  & \tablehead{1}{c}{b}{vol$_{50, 13}$~[pc$^{3}$]\tablenote{d$\le$50, V$\le$13}\\}
  & \tablehead{1}{c}{b}{vol$_{100, 15}$~[pc$^{3}$]\tablenote{d$\le$100, V$\le$15}\\}
  & \tablehead{1}{c}{b}{N$_{50, 13}^{\ast\ast}$\\}
  & \tablehead{1}{c}{b}{N$_{100, 15}^{\ddagger}$\\} \\
\hline
 8 - 9  & 1.698 $\times$ 10$^{-6}$ & 79.4 & 199.5 & 5.236 $\times$ 10$^{5}$ & 4.189 $\times$ 10$^{6}$ &  1 &   7 \\
 9 - 10 & 1.076 $\times$ 10$^{-5}$ & 50.1 & 125.9 & 5.236 $\times$ 10$^{5}$ & 4.189 $\times$ 10$^{6}$ &  6 &  45 \\
10 - 11 & 1.380 $\times$ 10$^{-4}$ & 31.6 &  79.4 & 1.325 $\times$ 10$^{5}$ & 2.099 $\times$ 10$^{6}$ & 18 & 290 \\
11 - 12 & 5.222 $\times$ 10$^{-4}$ & 20.0 &  50.1 & 3.327 $\times$ 10$^{4}$ & 5.273 $\times$ 10$^{5}$ & 17 & 275 \\
12 - 13 & 7.087 $\times$ 10$^{-4}$ & 12.6 &  31.6 & 8.358 $\times$ 10$^{3}$ & 1.325 $\times$ 10$^{5}$ &  6 &  94 \\
13 - 14 & 1.044 $\times$ 10$^{-3}$ &  7.9 &  20.0 & 2.099 $\times$ 10$^{3}$ & 3.327 $\times$ 10$^{4}$ &  2 &  35 \\
14 - 15 & 1.269 $\times$ 10$^{-3}$ &  5.0 &  12.6 & 5.273 $\times$ 10$^{2}$ & 8.358 $\times$ 10$^{3}$ &  1 &  11 \\
15 - 16 & 1.119 $\times$ 10$^{-3}$ &  3.2 &   7.9 & 1.325 $\times$ 10$^{2}$ & 2.099 $\times$ 10$^{3}$ &  0 &   2 \\
16 - 17 & 7.460 $\times$ 10$^{-5}$ &  2.0 &   5.0 & 3.327 $\times$ 10$^{1}$ & 5.273 $\times$ 10$^{2}$ &  0 &   0 \\
\hline
{\bf Tot} & 4.888 $\times$ 10$^{-3}$ &      &       &                         &                         & 51 & 759 \\
\hline
\end{tabular}
\caption{Statistics of nearby white dwarfs}
\label{tab:a}
\end{table}

\begin{figure}[!h]
\hspace{-11mm}
\includegraphics[height=.727\textheight]{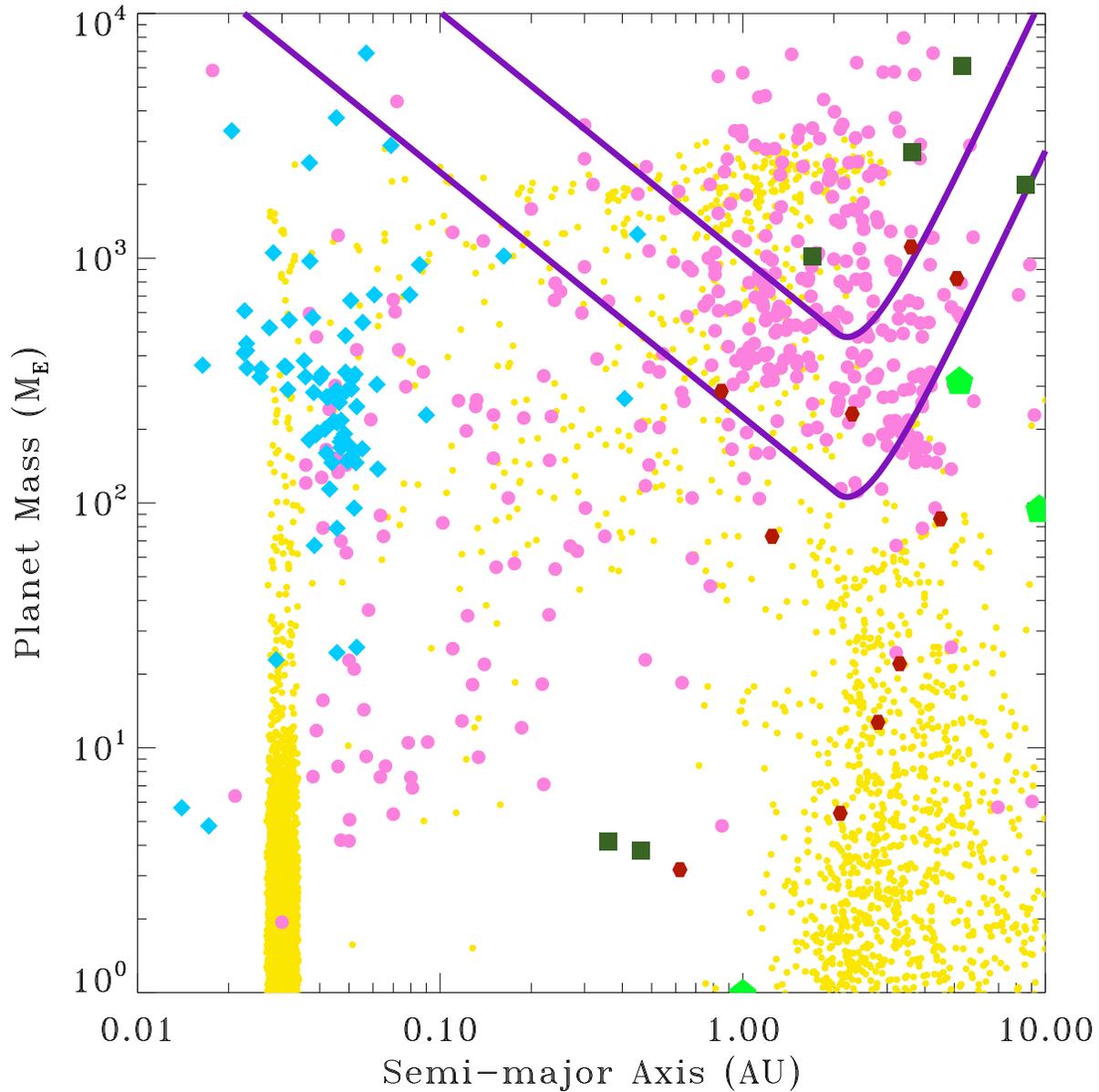}
\caption{Exoplanet discovery space for the GAIA mission based on double-blind 
tests \cite{Casertano+2010}.
Detectability curves are defined on the basis of a 3$\sigma$ criterion for 
signal detection. 
The upper and lower curves are for GAIA astrometry with 
$\sigma_A$ = 15 $\mu$as (where $\sigma_A$ is the single-measurement 
astrometric precision), assuming a 0.59-M$_{\odot}$ WD primary at 100 pc 
(V$<$15) and 50 pc (V$<$13), respectively. 
Survey duration is set to 5 yr. (Pink) dots indicate the inventory of 
Doppler-detected exoplanets as of May 2010. 
Transiting systems are shown as (light-blue) filled diamonds, while the (red) 
hexagons and the (dark-green) squares are planets detected by microlensing or 
timing respectively.
When the inclination of the system is not known, in particular for the 
(pink) dots and the (dark-green) squares, we used the minimum mass. 
Solar System planets are shown as (light-green) pentagons. 
The small (yellow) dots represent a theoretical distribution of masses and 
final orbital semi-major axes from \cite{IdaLin2008}.
The colors are visible only in the electronic version of the paper.}
\end{figure}

\section{Discussion}

As shown in Table 1 and Figure 1, GAIA will be able to verify the presence of 
Jovian planets (down to 0.7 M$_{Jup}$ or $\sim$200 M$_{Earth}$) around $\sim$50
bright (V$<$13) white dwarfs within 50 pc from the Sun.
For magnitudes fainter than V$\simeq$13, the GAIA performances start to 
decrease \cite{Lindegren2010}.
However, considering e.g. a larger sample of about 750 white dwarfs with V$<$15
and d$<$100 pc, GAIA will still be able to detect giant planets and brown 
dwarfs with masses larger than 2 M$_{Jup}$.

Considering that, up to now, the statistics of WD planets is close to zero, 
the contribution of GAIA will be essential to study the ancient evolved planetary systems around white dwarfs. 
This field is of particular interest since UV and optical spectroscopy has 
revealed heavy elements around $\sim$25\% of cool DA and $\sim$30\% of 
13.5-19.5 kK DB ``polluted'' white dwarfs (Zuckerman et al. 2010 and ref.
therein). 
These high-Z elements likely originate from asteroids or more massive rocky 
bodies that are tidally disrupted into a disk before being accreted onto the 
white dwarf.
Increasing evidence from IR studies (in particular from Spitzer) suggests that 
a fraction of these polluted white dwarfs are indeed surrounded by debris disks
(Farihi et al. 2010 and ref. therein). 
Following our projection of the WD local sample statistics, $\sim$13\% of
the stars reported in Table 1 are expected to show heavy elements (belonging to
the DAZ, DBZ, DQZ and DZ classes) and will give the possibility to study the
correlation between Z abundance and presence of planets.

Another interesting topic concerning WD planets is the possibility that 2nd 
(or even 3d) generation planets are formed with the material lost by the star
during its evolution, in particular during the red giant branch, asymptotic 
giant branch and planetary nebula ejection.
While it is likely that close-in 1st generation planets are engulfed and/or 
destroyed during these critical phases (Villaver, these proceedings),
2nd or 3d generation planets could form, in particular around old evolved
binary systems (Perets, these proc.).
The WD planets that will be discovered by GAIA will help to study these aspects as well.




\bibliographystyle{aipproc}   



\begin{thebibliography}{9}

\bibitem[(Casertano et al. 2010)]{Casertano+2010}
S.~Casertano, M.~Lattanzi, A.~Sozzetti et al., \emph{Astronomy \& Astrophysics} 
\textbf{482}, 699--729 (2008).

\bibitem[(Desidera \& Barbieri)]{Desidera+2007}
S.~Desidera, M.~Barbieri, \emph{Astronomy \& Astrophysics} \textbf{462},
345--353 (2007).

\bibitem[(Farihi et al. 2010)]{Farihi+2010} 
J.~Farihi, M.~Jura, J.-E.~Lee, B.~Zuckerman, \emph{Astroph.~Journal} 
\textbf{714}, 1386--1397 (2010).


\bibitem[(Holberg \& Bergeron 2006)]{HolbergBergeron2006}
J.~Holberg and P.~Bergeron, \emph{Astron.~Journal} \textbf{132}, 1221--1233 
(2006).

\bibitem[Holberg et al. (2008)]{Holberg+2008}
J.~B.~Holberg, E.~M.~Sion, T.~Oswalt et al., \emph{Astron.~Journal} 
\textbf{135}, 1225--1238 (2008).

\bibitem[Ida \& Lin (2008)]{IdaLin2008} 
S.~Ida and D.~N.~C.~Lin, \emph{Astroph.~Journal} \textbf{685}, 584--595 (2008).

\bibitem[(Jordan 2007)]{Jordan2007}
S.~Jordan, \emph{ASP Conf. Ser.} \textbf{372}, 139--144 (2007).


\bibitem[(Lindegren 2010)]{Lindegren2010}
L.~Lindegren, \emph{Proc. of the IAU} \textbf{261}, 296--305 (2010).

\bibitem[(Mullally et al. 2008)]{Mullally+2008}
F.~Mullally, D.~E.~Winget, S.~Degennaro et al., \emph{Astroph.~Journal} 
\textbf{676}, 573--583 (2008).

\bibitem[(Mullally et al. 2009)]{Mullally+2009}
F.~Mullally, W.~T.~Reach, S.~Degennaro, A.~Burrows, \emph{Astroph.~Journal} 
\textbf{694}, 327--331 (2009).





\bibitem[(Sigurdsson et al. 2003)]{Sigurdsson+2003}
S.~Sigurdsson, H.~B.~Richer, B.~M.~Hansen, I.~H.~Stairs, S.~E.~Thorsett,
\emph{Science} \textbf{301}, 193--196 (2003).


\bibitem[(Sion et al. 2009)]{Sion+2009}
E.~M.~Sion, J.~B.~Holberg, T.~D.~Oswalt, G.~P.~McCook, R.~Wasatonic,
\emph{Astron.~Journal} \textbf{138}, 1681--1689 (2009).

\bibitem[(Thorsett et al. 1999)]{Thorsett+1999}
S.~E.~Thorsett, Z.~Arzoumanian, F.~Camilo, A.~G.~Lyne, \emph{Astroph.~Journal}
\textbf{523}, 763--770 (1999).

\bibitem[(Wickramasinghe et al. 2010)]{Wickramasinghe+2010} 
D.~T.~Wickramasinghe, J.~Farihi, C.~A.~Tout, L.~Ferrario, R.~J.~Stancliffe, \emph{MNRAS} \textbf{404}, 1984--1991 (2010).

\bibitem[(Zuckerman et al. 2010)]{Zuckerman+2010}
B.~Zuckerman, C.~Melis, B.~Klein, D.~Koester, M.~Jura, \emph{Astroph.~Journal},
in press, arXiv:1007.2252v1

%
%
%

\end{thebibliography}





\end{document}